\documentclass[12pt]{article} \topmargin=-1.5cm \oddsidemargin=0pt \textheight=24cm \textwidth=16cm 
\usepackage[T2A]{fontenc}
\usepackage[utf8]{inputenc}
\usepackage[english]{babel}
\usepackage{amsmath}
\usepackage{amsfonts}
\usepackage{amssymb}
\usepackage{cite}
\usepackage{graphicx}
\usepackage{float}
\usepackage{color}

\title{Diffraction of large-number whispering gallery mode by boundary straightening with jump of curvature}
\author{E.A. Zlobina}
\date{}

\begin{document}

\maketitle

\begin{abstract}
Diffraction of a high-frequency large-number whispering gallery mode is studied, which runs along a concave curve turning to a straight line.
At the point of straitening, the curvature of the boundary suffers a jump.
The parabolic equation method is developed in the problem, and asymptotic formulas are presented for all waves arising in the vicinity of the non-smoothness point of the boundary.
The ``ray skeleton'' of the wavefield is investigated in detail.
\end{abstract}

\textit{Key words:} high-frequency asymptotics, diffraction by non-smooth obstacles, Helmholtz equation, parabolic equation method

\textit{MSC:} 35J25, 35L05

\section{Introduction}
Active investigation of high-frequency problems of diffraction by boundaries with jumping curvature,\footnote{Weaker singularities of curvature were also considered, see, e.g. \cite{AA}.} conducted since the 1960s (e.g., \cite{Wes,KamKel,Pop71,RogKis,KirPhi95,KirPhi97,KirPhi98,ZloKisWM,Zlo,ZloKisRE,ZKWM23,ZKAkZh}), was motivated by both possible technical applications and scientific curiosity.
At the moment, only the case of non-tangential incidence has been fully studied, in which the signs of curvature on different sides of the singular point of boundary have little effect on the analysis.
Conversely, in the case of tangential incidence, the signs of curvature are of fundamental importance. 
The arising problems differed widely, and their study requires more sophisticated approaches than in the non-tangential case.

In the articles devoted to tangential incidence, of which \cite{Pop71} and \cite{KirPhi95,KirPhi97,KirPhi98} should be noted, 
the wavefield was assumed to depend on time by the harmonic law $e^{-i\omega t}$ and was described by the Helmholtz equation
\begin{equation}\label{He}
    u_{xx} + u_{yy} + k^2 u = 0
\end{equation}
with the wavenumber $k \gg 1$ playing a role of large parameter.
The works did not pretend to study in detail the field in the vicinity of non-smoothness point. 
The main attention was paid to deriving expressions for surface currents and cylindrical diffracted waves diverging from the point of non-smoothness
\begin{gather}
u^{\text{dif}} = A(\varphi; k) \frac{e^{ikr}}{\sqrt{kr}}\left(1+\mathcal{E}(\varphi, r; k) \right), \label{cyl}\\
kr \gg 1, \label{kr}
\end{gather}
which was predicted by the Geometrical Theory of Diffraction \cite{BorKin}.
Here, $r$ and $\varphi$ are the classical polar coordinates, see Fig. \ref{f1},
\begin{equation}\label{pol}
x = r\cos\varphi, \quad y = r\sin\varphi, \quad 0 \leq r, \,\, -\pi \leq \varphi < \pi,
\end{equation}
$A$ is a diffraction coefficient, and $\mathcal{E}$  is a correction term which is not typically uniform in angle.
Transition zones which surround limit rays where diffraction coefficients have singularities were not considered in detail.

A complete asymptotic investigation of the field in the vicinity of non-smoothness point was undertaken in \cite{ZKWM23} and \cite{ZKAkZh}. 
In the work \cite{ZKWM23}, which continues the study \cite{Pop71}, the problem of diffraction of a plane wave running along a straight line passing into a convex curve (for the Neumann condition) was considered. 
In \cite{ZKAkZh}, as well as in \cite{KirPhi95}, diffraction of a whispering gallery mode at a point of jumply straightening of the boundary was studied, see Fig. \ref{f1}, and the mode number was small, that is, the number of transverse oscillations of the incident wave was not large. 
The problems were investigated within the framework of the parabolic equation method \cite{Foc,BabKir}, which is a boundary layer approach originated in V.A. Fock's works. 
A fundamental assumption for the method is that in a boundary layer the total wavefield has a general rapid oscillation in the longitudinal direction, and changes more slowly in the transverse direction.

This work continues the research undertaken in \cite{ZKAkZh}. 
Now, a much more complicated case is considered, when the number of transverse oscillations of the incident whispering gallery mode is large, but the parabolic equation method is still applicable. 
With its help, we study the wavefield in a small neighborhood of the point of non-smoothness. 
Then, matching the obtained expressions with a cylindrical wave, we find the diffraction coefficient, see \eqref{cyl}. 
It turns out that asymptotic formulas for the field in the case of incidence of the large-number whispering gallery mode are crucially different from the expressions obtained in the case of incidence of the small-number mode.
As in the works \cite{LMMP, Berry} devoted to a somewhat similar problem of diffraction at a inflection point, we pay a lot of attention to the ray structure of the field.

\section{Problem statement}
\begin{figure}[H]
    \noindent\centering{\includegraphics[width=0.5\textwidth]{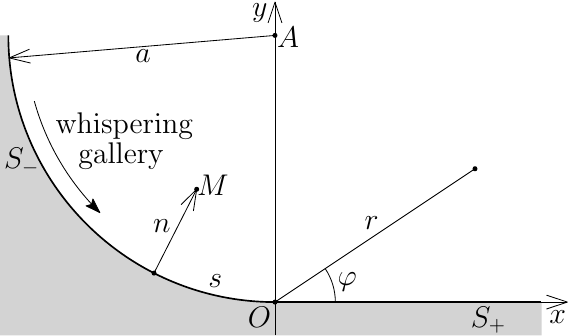}}
    \caption{Geometry of the problem}\label{f1}
\end{figure}

The wavefield $u$ is considered over the boundary $S = S_-\cup S_+$ which is an arc $S_-$ of circle of radius $a>0$, passing into a tangent straight line $S_+$ at the point $O$ (see Fig. \ref{f1}), so that the curvature of $S$ jumps from $1/a$ to $0$. 
There, $u$ satisfies the Helmholtz equation \eqref{He}, and the Neumann condition holds on the boundary\footnote{The case of the Dirichlet condition is treated similarly, and we will not dwell on it.}
\begin{equation}\label{cond}
    \left. u_n \right|_S = 0,
\end{equation}
where $u_n$ is a derivative along the normal to $S$. 
The wavefield vanishes far from the boundary: $u \to 0$.

We characterize the observation point $M = (x, y)$ positioned near $O$ by the distance $n \geq 0$ to the boundary along the normal and the arc length $s$ measured from $O$ (see Fig. \ref{f1}), which is negative as $x<0$ and positive as $x>0$ (there, $s=x$ and $n=y$).
Note that the mapping $(x, y) \to (s,n)$ is not smooth.

It is convenient to introduce a Fock-type dimensionless large parameter (arising in various problems with tangential incidence \cite{Foc,BabKir,Pop71,ZKAkZh})
\begin{equation}\label{M}
    \mathfrak{m} = \left(ka/2\right)^{\frac{1}{3}}, \quad \mathfrak{m}\gg 1,
\end{equation}
and pass to dimensionless stretched variables
\begin{equation}\label{xystr}
    \sigma= \mathfrak{m} s/a, \quad \nu = 2\mathfrak{m}^2 n/a.
\end{equation}

Along the concave part of the boundary, an incident whispering gallery mode $u^{\text{inc}}$ runs to the point $O$. 
In a small neighborhood of $O$, where
\begin{equation}\label{tube}
    \nu \ll \mathfrak{m}, \quad \mathfrak{t} \nu \ll \mathfrak{m}^2, \quad \mathfrak{t}^2 \sigma\ll \mathfrak{m}^2,
\end{equation}
$u^{\text{inc}}$ has a form \cite{BabKir,BabBul}
\begin{equation}\label{wgas}
    u^{\text{inc}} \approx e^{iks} U^{\text{inc}}, \quad U^{\text{inc}}(\sigma,\nu) = e^{-i\mathfrak{t} \sigma} v\left(\nu - \mathfrak{t} \right).
\end{equation}
Here, $v(z) = \sqrt{\pi}\mathrm{Ai}(z)$ is the Airy function (in Fock's notation \cite{Foc,BabKir}) satisfying the Airy equation $v''(z) = zv(z)$, with the following asymptotics:
\begin{align}
  v(z)=&\frac{1}{2z^{\frac{1}{4}}} e^{-\frac{2}{3} z^{\frac{3}{2}}} \left(1 + O\left(\frac{1}{z^{\frac{3}{2}}}\right)\right), \quad z \to +\infty, \label{vas}\\
  v(z)=&\frac{1}{(-z)^{\frac{1}{4}}}\sin\left(\frac{2}{3}(-z)^{\frac{3}{2}} + i\frac{\pi}{4}\right) \left(1 + O\left(\frac{1}{z^3}\right)\right) \\
  - &\frac{5}{24(-z)^{\frac{7}{4}}}\cos\left(\frac{2}{3}(-z)^{\frac{3}{2}} + i\frac{\pi}{4}\right) \left(1 + O\left(\frac{1}{z^3}\right)\right)
  \quad z \to -\infty \label{vas1},
\end{align}
and $(-\mathfrak{t})$ is one of zeros of its derivative, $v'(-\mathfrak{t})=0$, which are real, negative and simple (e.g., \cite{BabBul}).
In contrast to \cite{KirPhi95} and \cite{ZKAkZh}, here, $\mathfrak{t}$ is assumed to be large:
\begin{equation}\label{tgg}
    \mathfrak{t} \gg 1.
\end{equation}
Thus, in addition to the classical Fock-type parameter $\mathfrak{m}$ \eqref{M}, there is the second large parameter $\mathfrak{t}$ in the problem. 
The following relation 
\begin{equation}\label{tM}
    \mathfrak{t} \ll \mathfrak{m}^\frac{4}{5}
\end{equation}
is assumed to be fulfilled, importance of which will be clarified during a geometrical analysis of the problem (see the remark after \eqref{before}).

Not too far from the boundary, namely in the area where $\mathfrak{t} - \nu \gg 1$, the whispering gallery mode \eqref{wgas} oscillates along the normal to $S_-$, see \eqref{vas1}. 
With the growth of $z$, the function $v(z)$ and its derivatives of all orders decay exponentially, so the incident wave is localized in the vicinity 
\begin{equation}\label{sloj}
    0 \leq \nu < \mathfrak{t} + O(1)
\end{equation}
of concave part of the boundary $S_-$.

In this work, we aim at an asymptotic description of all waves arising to the right of the point of non-smoothness $O$ in its small neighborhood.
The research plan is as follows.
First, a detailed geometrical analysis of the problem will be carried out, which allows to characterize areas with different wavefield structures. 
After that, we proceed to an analytical investigation.
The wavefield in a small neighborhood of the point $O$ will be studied within the framework of the parabolic equation method.
The area of its applicability, as in the classical case \cite{Foc,BabKir}, is limited to the vicinity of the point $O$, where $s\ll 1$ and $n\ll 1$ (correspondingly, $x\ll 1$ and $y\ll 1$), but $\sigma$ and $\nu$ can be large. 
We will assume that inequalities 
\begin{equation}\label{obl}
    |\sigma| \ll \mathfrak{m}^{\frac{2}{5}}, \quad \nu \ll \mathfrak{m}
\end{equation}
hold true, which obviously agree with \eqref{tube}.
The restriction on $\nu$ follows directly from \eqref{tube}, and the need for such a restriction on $\sigma$ is explained after \eqref{before}.
The same explicitly solvable scattering problem for the parabolic equation arises as in \cite{ZKAkZh}, but analysis of its solution is complicated now by the presence of the second large parameter \eqref{tgg}.

\section{Geometrical considerations}
\subsection{Qualitative discussion of ray structure of wavefield}
\begin{figure}[H]
    \noindent\centering{\includegraphics[width=0.7\textwidth]{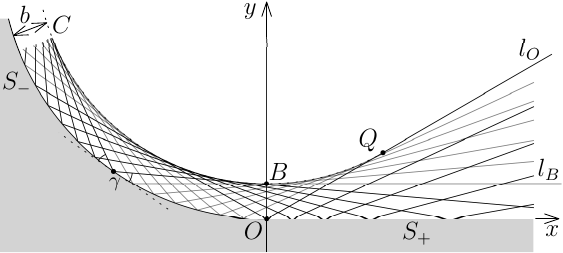}}
    \caption{Ray structure of wavefield (diffracted wave is omitted)}\label{f2}
\end{figure}

Consider the wavefield to the left of $y$ axis.
It is well known \cite{BabBul} that there is a congruence of rays associated with the whispering gallery mode, which reflect from the concave part of the boundary $S_-$, see Fig. \ref{f2}.
Since $S_-$ is an arc of circle, all the rays arrive to the boundary with the same grazing angle which we denote by $\gamma$. 
The rays corresponding to the incident wave \eqref{wgas} are localized in the boundary layer of width $b$ between $S_-$ and the caustic $C$ (see Fig. \ref{f2}). 
The latter is an arc of the circle
\begin{equation}\label{caus1}
    (y-a)^2 + x^2 = (a-b)^2
\end{equation}
of radius $a-b$ centred in the point $A$ (see Fig. \ref{f1}).
Bearing in mind limitation \eqref{sloj}, we find 
\begin{equation}\label{tb}
    \mathfrak{t} = 2\mathfrak{m}^2 b/a + O(1),
\end{equation}
see \eqref{M}, \eqref{xystr}. 
From \eqref{tM} and \eqref{tb}, the smallness of $b$ follows, $b/a \ll 1$, that implies the smallness of grazing angle: $\gamma \ll 1$.
An elementary geometrical consideration gives $b = a(1 - \cos\gamma)$, from where, taking into account \eqref{tb} and the smallness of $\gamma$, we derive the following important relations
\begin{equation}\label{b}
    b = a\gamma^2/2 \left(1+O\left(\gamma^2\right)\right),
\end{equation}
and
\begin{equation}\label{gamma}
    \mathfrak{m}\gamma = \sqrt{\mathfrak{t}}\left(1 + O\left(\mathfrak{t}/\mathfrak{m}^2 \right)\right).
\end{equation}

The ray structure described above is kept up to the $y$ axis and collapses to the right of it.

Consider the wavefield to the right of $y$ axis. 
There, rays split out into two families $\ell_1$ and $\ell_2$. 
The first ones, shown in Fig. \ref{f2} in black, at the point of intersection with the $y$ axis are directed downward, the second ones, shown in gray, are directed upward or horizontally.
The waves associated with the ray families $\ell_1$ and $\ell_2$ are denoted by $u_1$ and $u_2$, respectively.

The rays $\ell_1$ pass the caustic $C$ to the left of the point $O$ and reflect for the last time from the straight part $S_+$ of the boundary.
Before the reflection, they go below the horizontal ray $l_B$ crossing the $y$ axis at the point $B=(0, b)$, see Fig. \ref{f2}.
After the reflection, they propagate below the ray $l_O$ reflected at the point $O$, which is obviously characterized by the equation
\begin{equation}\label{lO1}
    y = x\tan\gamma = x \gamma \left(1+O\left(\gamma^2\right)\right). 
\end{equation}
Continuations of the reflected segments of rays below the boundary $S$ are symmetrical to the sections before the reflection, and therefore have an envelope symmetrical to the caustic $C$ relative to the $x$ axis.

The rays $\ell_2$ reflect for the last time from the concave part $S_-$ of the boundary. 
In some area to the right of $O$, they retain the structure of the incident wave, and the caustic $C$ continues up to the point $Q$ of its touch with the ray $l_O$, see Fig. \ref{f2}. 
From equation \eqref{caus1} of the caustic and equation \eqref{lO1} of the ray $l_O$, it immediately follows that $Q$ has coordinates
\begin{equation}\label{Qxy}
    x_Q = a\sin\gamma\cos\gamma = a\gamma \left(1+O\left(\gamma^2\right)\right), \quad y = a\sin^2\gamma = a\gamma^2 \left(1+O\left(\gamma^2\right)\right).
\end{equation}
Before passing the caustic, the rays $\ell_2$ propagate above the ray $l_O$. 
After passing the caustic, they go to infinity between the rays $l_B$ and $l_O$.
Thus, the ray $l_O$ (to the right of the point $Q$) is the shadow boundary both for the waves $u_1$ and $u_2$.
We call it the \textit{limit ray}.

Rays corresponding to the diffracted wave diverging from the point $O$ are omitted in Fig. \ref{f2}. 

Now we proceed to a detailed geometrical investigation of the rays of families $\ell_1$ and $\ell_2$. 
The results will be used in the analytical study of the problem.

\subsection{Rays of families $\ell_1$ and $\ell_2$}
\begin{figure}[H]
    \noindent\centering{\includegraphics[width=0.7\textwidth]{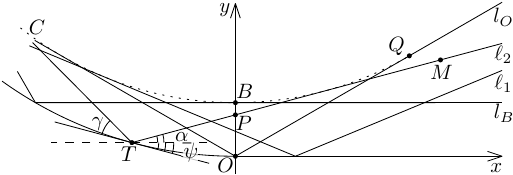}}
    \caption{Rays of families $\ell_1$ and $\ell_2$, limit ray $l_O$ and horizontal ray $l_B$}\label{f3}
\end{figure}

Let the observation point $M=(x,y)$ lie in a small neighborhood of $O$. 
Consider a ray $l$ coming to $M$ immediately after its reflection from the concave part of the boundary $S_-$ (i.e. $l$ is either a ray of the family $\ell_2$ or a ray of the family $\ell_1$ before its reflection from the straight part of the boundary $S_+$). 
The case when $l$ comes to $M$ after reflection from $S_+$ will be considered later.
We analytically describe position of the point $P=(0, h)$ of the intersection of $l$ with the $y$ axis and position of the point $T = (\xi, \eta)$ of the last reflection from $S_-$, see Fig. \ref{f3}.

\textbf{1.} Since the point $T$ is obviously close to $O$, its Cartesian coordinates $\xi$ and $\eta$ are expressed in terms of the corresponding arc length $\hat{s}$ (which is negative to the left of $O$):
\begin{equation}\label{xieta}
    \xi = \hat{s} + O\left(\hat{s}^3/a^2 \right), \quad \eta = \hat{s}^2/2a + O\left(\hat{s}^4/a^3 \right).
\end{equation}
The tangent to the boundary at the point $T$ forms an angle  
\begin{equation}\label{psi}
    \psi = -\hat{s}/a.
\end{equation}
with the horizontal line. 
The angle of inclination $\alpha$ of the ray $l$ is related to the $\gamma$ and $\psi$ as follows: 
\begin{equation}\label{alpha}
    \alpha = \gamma -\psi.
\end{equation}
Note that $\alpha <0$ (respectively, $\hat{s} < -a\gamma$) for all the rays $\ell_1$ and $\alpha\geq 0$ (respectively, $\hat{s} \geq -a\gamma$) for all the rays $\ell_2$.
For the ordinate of the point $P=(0, h)$, the equality 
\begin{equation}\label{eq1}
    h - \eta \approx -\xi \alpha
\end{equation}
is valid, where the smallness of $\alpha$ is used. 
Here and further, we omit smaller terms with respect to parameters $1/\mathfrak{m}$ and $\mathfrak{t}/\mathfrak{m}^2$ (see \eqref{M} and \eqref{tM}).
Substituting \eqref{xieta} into \eqref{eq1} and taking into account the smallness of $\gamma$ and $\hat{s}$, we get
\begin{equation}\label{h}
    h \approx -\hat{s}^2/2a - \hat{s}\gamma.
\end{equation}
Conditions $h \geq 0$ and $\hat{s} < 0$ yield a limitation
\begin{equation}\label{ogr}
    \hat{s} \geq -2a\gamma.
\end{equation}
Taking into account \eqref{psi} and \eqref{alpha}, an interesting relation follows from \eqref{h}:
\begin{equation}\label{b-h}
    \sqrt{b - h} \approx \sqrt{a \left(\gamma^2 + 2\gamma\hat{s}/a + (\hat{s}/a)^2\right)/2} = \sqrt{a(\gamma - \psi)^2/2} = |\alpha|\sqrt{a/2},
\end{equation}
which links the angle of inclination of the ray $l$ to the distance between the points of intersection of the ray $l$ and the horizontal ray $l_B$ with the $y$ axis (see Fig. \ref{f3}). 
In the stretched coordinates, \eqref{b-h} has the form
\begin{equation}\label{t-p}
    \sqrt{\mathfrak{t}-p} \approx \mathfrak{m}|\alpha|,
\end{equation}
cf. \eqref{gamma}, where 
\begin{equation}\label{p}
    p = 2\mathfrak{m}^2 h/a.
\end{equation}

\textbf{2.} Condition, ensuring that the points $T=(\xi, \eta)$, $P=(0, h)$ and $M=(x, y)$ lie on the same straight line, reads
\begin{equation}\label{line}
    (y - h)/x = (\eta - h)/\xi.
\end{equation}
Substituting expressions for the coordinates of the contour's point $T$ \eqref{xieta} and the point $P$ \eqref{h} into \eqref{line}, we arrive at the quadratic equation for $\hat{s}$:
\begin{equation}\label{quadr}
    \hat{s}^2 + 2\hat{s} (a\gamma - x) + 2a(y - \gamma x) \approx 0.
\end{equation}
Its discriminant $D$, accounting relation \eqref{b}, is
\begin{equation*}
    D = 4((a\gamma - x)^2 - 2a(y - \gamma x)) = 8a \widetilde{n}.
\end{equation*}
The quantity
\begin{equation}\label{dist_c}
    \widetilde{n} = b + x^2/2a - y
\end{equation}
equals, in the main approximation, the distance from the observation point $M$ to the caustic $C$, see \eqref{caus1}. 
$D$ is positive when $M$ lies below $C$, vanishes when $M$ is the caustic point, and is negative when $M$ lies above $C$.\footnote{This agrees with the fact that the rays of the families $\ell_1$ and $\ell_2$ lie below the caustic.} 
We are interested in negative values of $\hat{s}$.
It is easy to show that for $x>a\gamma$ and $y\leq\gamma x$, i.e. for points $M$ lying to the right of $Q$ and below the limit ray $l_O$, see \eqref{lO1} and \eqref{Qxy},  equation \eqref{quadr} has one negative root
\begin{equation}\label{s1}
    \hat{s}_1 \approx x - a\gamma - \sqrt{2a\widetilde{n}}.
\end{equation}
If $x<a\gamma$ and $y>\gamma x$, i.e. $M$ lies to the left of $Q$ and above $l_O$, then the second root 
\begin{equation}\label{s2}
    \hat{s}_2 \approx x - a\gamma + \sqrt{2a\widetilde{n}}
\end{equation}
becomes negative.
In the case when $M$ does not lie in the areas specified above, equation \eqref{quadr} has no negative solutions.
For $y<b$, i.e. for $M$ lying below $l_B$ (see Fig. \ref{f3}), $\hat{s}_1 < -a\gamma$ describes the position of the reflection point $T$ of the ray of the family $\ell_1$. 
If $y\geq b$, i.e. $M$ is located above $l_B$, then $\hat{s}_1\geq -a\gamma$ characterizes the point $T$ for the ray of the family $\ell_2$, which came to $M$ after passing the caustic.   
Expression \eqref{s2} takes values $\hat{s}_2> -a\gamma$ and describes the position of the point $T$ if the ray of the family $\ell_2$ came to $M$ before passing caustic.
Note that for $M$ lying on the caustic, $\widetilde{n}=0$ and $\hat{s}_1 = \hat{s}_2$.

\textbf{3.} Position of the point $T$, at which the ray, arriving to the point $M$ after reflection from the straight part of the boundary, was last reflected from $S_-$, is determined quite similarly (the calculations differ from those above only by replacing $y$ with $-y$):
\begin{equation}\label{s3}
    \hat{s}_3 \approx x - a\gamma - \sqrt{2a(\widetilde{n} + 2y)},
\end{equation}
see \eqref{dist_c}. 
The quantity 
\begin{equation}\label{dist_c+}
    \widetilde{n} + 2y = b + x^2/2a + y
\end{equation}
characterizes the distance from the point $M'$, symmetric to the observation point $M$ relative to $S_+$, to the caustic $C$. 
Condition \eqref{ogr} implies the inequality $y\leq\gamma x$, i.e. the point $M$ must lie below the limit ray $l_O$.

\textbf{4.} We obtain the ordinate of the intersection point $P$ of the ray $l$ with the $y$ axis by substituting corresponding arc length \eqref{s1}, \eqref{s2}, \eqref{s3} into \eqref{h}. 
We give expressions for stretched ordinate \eqref{p}, since they are the ones we will encounter during analytical considerations.

1) $M$ lies on a ray of the family $\ell_1$ before its reflection from $S_+$ or on the ray of the family $\ell_2$ after passing the caustic:
\begin{equation}\label{p1}
    p_1 = p(\hat{s}_1) = \mathfrak{t} - \left(\sigma- \sqrt{\widetilde{\nu}}\right)^2;
\end{equation}

2) $M$ lies on a ray of the family $\ell_2$ before passing the caustic:
\begin{equation}\label{p2}
    p_2 = p(\hat{s}_2) = \mathfrak{t} - \left(\sigma+ \sqrt{\widetilde{\nu}}\right)^2;
\end{equation}

3) $M$ lies on a ray of the family $\ell_1$ after its reflection from $S_+$:
\begin{equation}\label{p3}
    p_3 = p(\hat{s}_3) = \mathfrak{t} - \left(\sigma- \sqrt{\widetilde{\nu} + 2\nu}\right)^2.
\end{equation}
Here,
\begin{equation}\label{kap}
    \widetilde{\nu} = \mathfrak{t} + \sigma^2 -\nu
\end{equation}
is the stretched distance to the caustic, cf. \eqref{dist_c}. 

Now we proceed to the calculation of the eikonals (geometrical travel times) of the waves $u_1$ and $u_2$ corresponding to the ray families $\ell_1$ and $\ell_2$.
Eikonals will first arise in the study of the geometry of rays, but, as is usual in the high-frequency theory of diffraction \cite{BabBul,BorKin}, they will also appear in the analytical analysis of the problem.

\subsection{Eikonals of waves $u_1$ and $u_2$}
\begin{figure}[H]
    \noindent\centering{\includegraphics[width=0.4\textwidth]{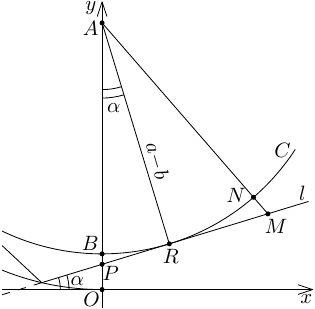}}
    \caption{Calculation of eikonals}\label{f4}
\end{figure}

\textbf{1.} Let the ray $l$ arrive to the observation point $M=(x,y)$ after passing the caustic $C$ as shown in Fig.~\ref{f4}.
As before, we denote by $P=(0, b)$ the point of its intersection with the ordinate axis, for $R$ the point of touch with caustic, and by $\alpha$ the angle of inclination, which can be either positive or negative. 
In Fig.~\ref{f4}, the case $\alpha > 0$ is shown.
It is clear that the angle $\angle BAR = |\alpha|$.
The value of eikonal at the point $M$, as is well known (see, for example, \cite{BabBul}), is expressed in terms of the arc lengths $\overset{\frown}{NR}$, $\overset{\frown}{NB}$ and the length of the segment $MR$: $\tau_+(M) = |\overset{\frown}{NB}| + |MR| - |\overset{\frown}{NR}|$. 
This transforms to
\begin{equation}\label{t+}
    \tau_+(M) = |MP| - (|PR| - |\overset{\frown}{BR}|).
\end{equation} 
Obviously, $|PR| = (a-b)\tan|\alpha|$ and $|\overset{\frown}{BR}| = (a-b)|\alpha|$, whence
\begin{equation}\label{razn}
    |PR| - |\overset{\frown}{BR}| = \frac{a-b}{3}|\alpha|^3\left(1+O\left(\alpha^2\right)\right) = \frac{a}{3}|\alpha|^3\left(1 + O\left(\frac{b}{a}\right) + O\left(\alpha^2\right)\right).
\end{equation}
Length of the segment $MP$ equals
\begin{equation}\label{len}
    |MP| = \sqrt{x^2 + (y-h)^2} = x + \frac{(y-h)^2}{2x} + O\left(\frac{(y-h)^4}{x^3}\right).
\end{equation}
Substituting \eqref{razn} and \eqref{len} into \eqref{t+}, passing to stretched coordinates \eqref{xystr} and using relation \eqref{t-p}, we come up with the expression for eikonal of the wave $u_2$ after the corresponding rays $\ell_2$ pass the caustic, and for eikonal of the wave $u_1$ before the corresponding rays $\ell_1$ reflect from $S_+$:
\begin{equation}\label{after}
    k\tau_+ (M) = kx + \frac{(p-\nu)^2}{4\sigma} + \frac{2}{3}(\mathfrak{t}-p)^{\frac{3}{2}} + O\left(\frac{(\nu-p)^4}{\mathfrak{m}^2 \sigma^3} \right) + O\left(\frac{(\mathfrak{t}-p)^{\frac{3}{2}}\mathfrak{t}}{\mathfrak{m}^2}\right).
\end{equation}
To derive the formula for eikonal of $u_1$ after the rays $\ell_1$ reflect from $S_+$, it is enough to replace $\nu$ by $-\nu$ in \eqref{after}. 

\textbf{2.} If the ray $l$ arrives at the point $M$ before passing the caustic, then the eikonal has the form $\tau_-(M) = |\overset{\frown}{NB}| - |MR| + |\overset{\frown}{NR}|$, see \cite{BabBul}.
By transformations similar to those done above, we find an expression for the eikonal of the wave $u_2$ before the corresponding rays $\ell_2$ pass the caustic:
\begin{equation}\label{before}
    k\tau_- (M) = kx + \frac{(p-\nu)^2}{4\sigma} - \frac{2}{3}(\mathfrak{t}-p)^{\frac{3}{2}} + O\left(\frac{(\nu-p)^4}{\mathfrak{m}^2 \sigma^3} \right) + O\left(\frac{(\mathfrak{t}-p)^{\frac{3}{2}}\mathfrak{t}}{\mathfrak{m}^2}\right).
\end{equation}
Note that inequalities \eqref{tM} and \eqref{obl} provide the smallness of correction terms in \eqref{after} and \eqref{before} (since $\nu - p = 2\sigma\left(\sigma\pm \sqrt{\widetilde{\nu}}\right)$, see \eqref{p1}, \eqref{p2}). 
Thus, it is clear that \eqref{tM} and \eqref{obl} can not be weaked.

\textbf{3.} Substituting the corresponding expressions for $p$, see \eqref{p1}, \eqref{p2}, \eqref{p3}, into \eqref{after} and \eqref{before}, we establish that
 
1) the eikonal of the wave $u_2$ after the corresponding rays pass the caustic, and eikonal of the wave $u_1$ before reflection of the corresponding rays from the straight part of the boundary have the form
\begin{equation}\label{ph1}
    \phi_1 \approx kx - \sigma(\mathfrak{t} - \nu) -\frac{2}{3}\sigma^3 + \frac{2}{3}\widetilde{\nu}^{\frac{3}{2}}; 
\end{equation}

2) the eikonal of the wave $u_2$ before the corresponding rays pass the caustic is 
\begin{equation}\label{ph2}
    \phi_2 \approx kx - \sigma(\mathfrak{t} - \nu) -\frac{2}{3}\sigma^3 - \frac{2}{3}\widetilde{\nu}^{\frac{3}{2}};
\end{equation}

3) the eikonal of the wave $u_1$ after reflection of the corresponding rays from straight part of the boundary has the form
\begin{equation}\label{ph3}
    \phi_3 \approx kx - \sigma(\mathfrak{t} + \nu) -\frac{2}{3}\sigma^3 + \frac{2}{3}(\widetilde{\nu} + 2\nu)^{\frac{3}{2}}.
\end{equation}
The quantity $\widetilde{\nu}$ is introduced in \eqref{kap}.
Near caustic \eqref{caus1} $\widetilde{\nu}$ is small and $\phi_1 \approx \phi_2$.

\textbf{4.} The eikonal of cylindrical diffracted wave \eqref{cyl} is
\begin{equation}\label{ph_d}
    kr = k\sqrt{x^2+y^2} = kx + k\frac{y^2}{2x} + O\left(k\frac{y^4}{x^3}\right) = kx + \frac{\nu^2}{4\sigma} + O\left(\frac{\nu^4}{\mathfrak{m}^2\sigma^3}\right).
\end{equation}
Near the limit ray $l_O$ \eqref{lO1}, the approximation
\begin{equation}\label{ph_d_as}
    kr \approx kx + \sigma\mathfrak{t} - 2\sigma\sqrt{\mathfrak{t}}\, \varepsilon+\sigma\varepsilon^2
\end{equation}
is valid, where 
\begin{equation}\label{eps}
    \varepsilon= \frac{2\sqrt{\mathfrak{t}}\,\sigma- \nu}{2\sigma} = \sqrt{\mathfrak{t}} - \frac{\nu}{2\sigma}
\end{equation} 
is small.
Note that $\varepsilon$ characterizes the proximity of the observation point to the limit ray $l_O$:
\begin{equation}\label{epsg}
    \varepsilon \approx \mathfrak{m} (\gamma - \varphi),
\end{equation}
see \eqref{xystr}. 
Here, $\varphi$ is the polar angle, see \eqref{pol} and Fig.~\ref{f1}.
Near $l_O$, the eikonal of the wave $u_1$ after the rays $\ell_1$ reflection from $S_+$ has the form 
\begin{equation}\label{ph3_as}
    \phi_3 \approx \frac{2}{3}\mathfrak{t}^{\frac{3}{2}} + kx + \sigma\mathfrak{t} - 2\sigma\sqrt{\mathfrak{t}}\, \varepsilon+ \frac{\sigma^2 \varepsilon^2}{\sigma+ \sqrt{\mathfrak{t}}}
\end{equation}
which differs from \eqref{ph_d_as} in the first two orders only by a constant summand.
The same is true for eikonal of the wave $u_2$:
\begin{equation}\label{ph12_as}
    \phi_1 = \phi_2 \approx -\frac{2}{3}\mathfrak{t}^{\frac{3}{2}} + kx + \sigma\mathfrak{t} - 2\sigma\sqrt{\mathfrak{t}}\, \varepsilon+ \frac{\sigma^2 \varepsilon^2}{\sigma- \sqrt{\mathfrak{t}}}.
\end{equation}

\textbf{5.} With the help of geometrical considerations (in essence, elementary ones) we find a lot about the ``ray skeleton'' of the wavefield.
Now we proceed to the analytical analysis of the problem \eqref{He}, \eqref{cond}, \eqref{wgas}, during which the field near the caustic and in the transition zone around the limit ray will be described by special functions.

\section{Problem for parabolic equation and its exact solution}
The main term of the asymptotics of total wavefield is sought in a form similar to incident wave \eqref{wgas}:
\begin{equation}\label{uas}
    u = e^{iks} U(\sigma,\nu).
\end{equation}
Recall that $s=x$ to the right of $O$.
Following \cite{Foc,BabKir}, we call $U$ the attenuation factor.

Substituting \eqref{wgas} and \eqref{uas} into the Helmholtz equation \eqref{He} and boundary condition \eqref{cond}, we obtain the parabolic equation with a non-smooth coefficient
\begin{equation}\label{parab}
    i U_{\sigma} + U_{\nu\nu} - \nu \theta(-\sigma) U = 0,
\end{equation}
and a homogeneous boundary condition
\begin{equation}\label{gran}
    \left. U_{\nu} \right|_{\nu = 0}= 0.
\end{equation}
Here, $\theta$ is the Heaviside function, $\theta(z)=0$ as $z \leq 0$ and $\theta(z)=1$ as $z > 0$.

An exact solution of \eqref{parab}, \eqref{gran} is derived in \cite{KirPhi95,ZKAkZh}.\footnote{In \cite{ZKAkZh}, in contrast to \cite{KirPhi95}, the detailed formulation of correct scattering problem for parabolic equation is given.} 
As $\sigma\leq 0$, it coincides with $U^{\text{inc}}$, and as $s>0$ it is given by the expression
\begin{equation}\label{U}
    U = \frac{e^{-i\frac{\pi}{4}}}{2\sqrt{\pi \sigma}} \int_{0}^{\infty} v(p -\mathfrak{t}) \left(e^{i\frac{(p-\nu)^2}{4\sigma}} + e^{i\frac{(p +\nu)^2}{4\sigma}} \right) dp.
\end{equation}

So far, the analytical study of the problem with the large $\mathfrak{t}$ is the same as the investigation of the case $\mathfrak{t}= O(1)$, see \cite{ZKAkZh}. 
However, the asymptotic analysis of expression \eqref{U} in the case of $\mathfrak{t}\gg 1$, which is carried out below, differs greatly from the considerations of \cite{ZKAkZh}, since not only quadratic exponents, but also the Airy function can rapidly oscillate.

\section{Asymptotic investigation of attenuation factor}
Since $\mathfrak{t}\gg 1$, the integration interval in \eqref{U} naturally splits out into three intervals where the integrand behaves differently.

1) On the interval 
\begin{equation}\label{L1}
    L_1 = \{\mathfrak{t} - p \gg 1\}
\end{equation}
the function $v(p-\mathfrak{t})$ is replaced by rapidly oscillating asymptotics \eqref{vas1};

2) on the interval 
\begin{equation}\label{L2}
    L_2 = \{|\mathfrak{t}-p|=O(1)\}
\end{equation}
the Airy function varies slowly;

3) on the interval 
\begin{equation}\label{L3}
    L_3 = \{p - \mathfrak{t} \gg 1\}
\end{equation}
the function $v(p-\mathfrak{t})$ decays exponentially, see \eqref{vas}.

We investigate the integrals $I_1$, $I_2$, $I_3$ over the corresponding intervals $L_1$, $L_2$, $L_3$ separately.
The integral $I_3$ is obviously negligible compared to $I_2$. 
The main difficulty is the analysis of the integral $I_1$.

\subsection{Investigation of integral $I_1$}
\textbf{1.} We replace the Airy function with its asymptotics \eqref{vas1}, and thus the integral $I_1$ becomes: 
\begin{equation*}
    I_1 = I_1^{++} + I_1^{+-} + I_1^{-+} + I_1^{--}, 
\end{equation*}
where
\begin{gather*}
    I_1^{-\pm} = \frac{1}{4\sqrt{\pi \sigma}} \int_{L_1} \frac{e^{if^{-\pm}(p)}}{(\mathfrak{t}-p)^{\frac{1}{4}}}  \left(1 + \frac{5i}{24(\mathfrak{t} - p)^{\frac{3}{2}}} + O\left(\frac{1}{(\mathfrak{t} - p)^{3}}\right)\right) dp,\\
    I_1^{+\pm} = -\frac{i}{4\sqrt{\pi \sigma}} \int_{L_1} \frac{e^{if^{+\pm}(p)}}{(\mathfrak{t}-p)^{\frac{1}{4}}}  \left(1 - \frac{5i}{24(\mathfrak{t} - p)^{\frac{3}{2}}} + O\left(\frac{1}{(\mathfrak{t} - p)^{3}}\right)\right) dp,
\end{gather*}
and the phases $f^{++}$, $f^{+-}$, $f^{-+}$ and $f^{--}$ of the exponentials are
\begin{equation}\label{fpmpm}
    f^{\pm\pm}(p) = \pm i\frac{2}{3}(\mathfrak{t} - p)^{\frac{3}{2}} + \frac{(p\pm\nu)^2}{4\sigma}.
\end{equation}
The first superscript corresponds to the sign before $(\mathfrak{t} - p)^{\frac{3}{2}}$ in the phase, and the second to the sign before $\nu$.
Note that expressions \eqref{fpmpm} resemble formulas \eqref{after} and \eqref{before} for the ray eikonals.

\textbf{2.} The asymptotics of each of the integrals $I_1^{\pm\pm}$ is given by contributions of the end point $p=0$ and the critical points of phases \eqref{fpmpm}.

The critical points of the phases are found from the equations
\begin{equation*}
    (f^{\pm\pm}(p))' = \mp \sqrt{\mathfrak{t}-p} + \frac{p\mp\nu}{2\sigma} = 0.
\end{equation*}
We obtain by a direct calculation that

1) the phase $f^{+-}(p)$ has one critical point $p_1 = \mathfrak{t} - \left(\sigma- \sqrt{\widetilde{\nu}}\right)^2$ as $\mathfrak{t}>\nu$ ($\widetilde{\nu}$ is introduced in \eqref{kap}).

2) the phase $f^{--}(p)$ has one critical point $p_2 = \mathfrak{t} - \left(\sigma+ \sqrt{\widetilde{\nu} }\right)^2$ as $\mathfrak{t}>\nu$, and $f^{--}(p)$ has also another critical point $p_1 = \mathfrak{t} - \left(\sigma- \sqrt{\widetilde{\nu}}\right)^2$ as $\mathfrak{t}<\nu$.

3) the phase $f^{++}(p)$ has one critical point $p_3 = \mathfrak{t} - \left(\sigma- \sqrt{\widetilde{\nu} + 2\nu}\right)^2$.

4) the phase $f^{-+}(p)$ has one critical point $p_4 = \mathfrak{t} - \left(\sigma+ \sqrt{\widetilde{\nu} + 2\nu}\right)^2$.
Note that $p_4$ is negative and large in modulus for any values of $\sigma$ and $\nu$. It does not contribute to the integral.

The expressions for the critical points $p_1$, $p_2$, $p_3$ coincide with expressions \eqref{p1}, \eqref{p2}, \eqref{p3}, respectively, for the points of intersection of the $y$ axis by rays of the families $\ell_1$ and $\ell_2$. 
The contributions of $p_1$, $p_2$, $p_3$ to the asymptotics of integrals $I_1^{\pm\pm}$, as we will see below, describe the contributions of the corresponding rays to the wavefield.

The exact formulation of the limitations on $\sigma$ and $\nu$, ensuring that $p_1$, $p_2$, $p_3$ lie on the interval $L_1$ \eqref{L1}, is rather cumbersome, and we omit it.
Note only that the inequalities $\mathfrak{t} - p_{1,2,3}\gg 1$ (due to the remarks above and relation \eqref{t-p}) can be interpreted as a condition that the rays corresponding to the critical points are not too close to the horizontal ray $l_B$, see Fig. \ref{f3}.

\textbf{3.} Here we list the limitations on the stretched coordinates $(\sigma,\nu)$ of observation points $M$ that guarantee the positivity of critical points, which are not difficult to obtain from the geometrical considerations carried out in section 3.2:

1) the values of $p_3$ are real for all $\sigma$ and $\nu$, while $p_1$ and $p_2$ are real only when $\nu \leq \mathfrak{t} + \sigma^2$, i.e., $M$ lies below the caustic.

2) $p_1 \geq 0$ as $\sigma\leq \sqrt{\mathfrak{t}}$ and $\nu \geq 2\sqrt{\mathfrak{t}}\,\sigma$ (i.e., $M$ is positioned to the left of $Q$ and below the caustic, but above the limit ray $l_O$), and as $\sigma\geq \sqrt{\mathfrak{t}}$ and $\nu \leq 2\sqrt{\mathfrak{t}}\,\sigma$ (i.e., $M$ is positioned to the right of $Q$ and below the limit ray $l_O$). 

3) $p_2 \geq 0$ as $\sigma\leq \sqrt{\mathfrak{t}}$ and $\nu \geq 2\sqrt{\mathfrak{t}}\,\sigma$ (i.e., $M$ lies to the left of $Q$ and below the caustic, but above the limit ray $l_O$). 

4) near the limit ray, where $\nu \approx 2\sqrt{\mathfrak{t}}\,\sigma$, the points $p_1$ and $p_2$ merge with the end point $p=0$: $p_1$ merges as $\sigma\geq \sqrt{\mathfrak{t}}$, and $p_2$ merges as $\sigma\leq \sqrt{\mathfrak{t}}$.
If $M$ lies near caustic, where $\widetilde{\nu}$ \eqref{kap} is small, then the points $p_1$ and $p_2$ merge.
if $M$ is near the point $Q$ where caustic and the limit ray touch, then $p_1$, $p_2$ and $p=0$ merge all together.

5) $p_3 \geq 0$ as $\nu \leq 2\sqrt{\mathfrak{t}}\,\sigma$ (i.e., $M$ is positioned below the limit ray $l_O$). 
As $\nu \approx 2\sqrt{\mathfrak{t}}\,\sigma$ (i.e., $M$ lies near the limit ray $l_O$) $p_3 \approx 0$.

We present a detailed analysis of the integral $I_1^{--}$, which is the most difficult.
The other integrals are investigated using the same techniques, and for them we give only the final results.

\subsubsection{Integral $I_1^{--}$}
Consider the integral
\begin{equation*}
    I_1^{--} = \frac{1}{4\sqrt{\pi \sigma}} \int_{L_1} \frac{e^{if^{--}(p)}}{(\mathfrak{t}-p)^{\frac{1}{4}}}  \left(1 + \frac{5i}{24(\mathfrak{t} - p)^{\frac{3}{2}}} + O\left(\frac{1}{(\mathfrak{t} - p)^{3}}\right)\right) dp.
\end{equation*}

\textbf{1.} Let us first turn to the case when critical points of the phase $p_1$ and $p_2$ are far from each other and from the end point $p = 0$.
Then the asymptotics of $I_1^{--}$ is given by the sum of these three points' contributions calculated by standard methods \cite{Fed}.

Integrating by parts, we derive the contribution of the end point:
\begin{equation}\label{mm_0l}
    (I_1^{--})_0 = \frac{e^{-i\frac{2}{3}\mathfrak{t}^{\frac{3}{2}} + i\frac{\nu^2}{4\sigma}}}{4\sqrt{\pi \sigma} \mathfrak{t}^{\frac{1}{4}}} \Bigg[\frac{i}{\varepsilon}\left(1 + \frac{5i}{24 \mathfrak{t}^{\frac{3}{2}}} + O\left(\frac{1}{\mathfrak{t}^3}\right)\right) - \frac{1}{4\mathfrak{t} \, \varepsilon^2}
    + \left(\frac{1}{\sigma} - \frac{1}{\sqrt{\mathfrak{t}}}\right) \frac{1}{2 \varepsilon^3} \left(1 + \delta \right) \Bigg].
\end{equation}
Here, $\delta$ denotes the correction terms,
\begin{multline}\label{de}
    \delta = O\left(\frac{1}{\sigma\varepsilon^2}\right) + O\left(\frac{1}{\sqrt{\mathfrak{t}}\,\varepsilon^2}\right) + O\left(\frac{1}{\mathfrak{t}^2}\right) \\
    = O\left(\frac{1}{kr(\gamma-\varphi)^2}\right) + O\left(\frac{1}{ka\gamma(\gamma - \varphi)^2}\right) + O\left(\frac{1}{\mathfrak{t}^2}\right),
\end{multline}
and $\varepsilon$ is introduced in \eqref{eps}.
The need to keep so many terms of expansion \eqref{mm_0l} will be clarified later, when summing the contributions of the end point $p=0$ to the integrals $I_1^{\pm\pm}$.

Smallness of correction terms \eqref{de} in \eqref{mm_0l} is guaranteed by conditions \eqref{tgg} and
\begin{equation}\label{att1}
    \sigma\varepsilon^2 \gg 1,\,\, \sqrt{\mathfrak{t}}\,\varepsilon^2 \gg 1 \quad \Leftrightarrow \quad kr(\gamma -\varphi)^2 \gg 1, \,\, ka\gamma (\gamma -\varphi)^2 \gg 1.
\end{equation}
It follows from \eqref{att1} that the observation point $M$ is positioned far from the limit ray $l_O$, see Fig. \ref{f2}.
The first inequality often arises in diffraction problems, see, e.g., \cite{BorKin}. 

To calculate the contributions of the critical points, we write out the second derivative of the phase:
\begin{equation}\label{sec}
    (f^{--}(p))'' = \frac{1}{2}\left(\frac{1}{\sigma} - \frac{1}{\sqrt{\mathfrak{t}-p}}\right).
\end{equation}
At the points $p_1$ and $p_2$, respectively, expression \eqref{sec} takes the following values 
\begin{equation*}
    (f^{--}(p_1))'' = -\frac{\sqrt{\widetilde{\nu}}}{2\sigma(\sigma- \sqrt{\widetilde{\nu}})}, \quad (f^{--}(p_2))'' =  \frac{\sqrt{\widetilde{\nu}}}{2\sigma(\sigma+ \sqrt{\widetilde{\nu}})},
\end{equation*}
where the quantity $\widetilde{\nu}$ is defined in \eqref{kap}. 
Recall that $p_1$ is the critical point of the integral $I_1^{--}$ for $\nu > \mathfrak{t}$ only, thus $(f^{--}(p_1))''<0$.
Note that when the observation point $M$ is close to the caustic, $\widetilde{\nu}$ is small, and the values of the second derivative of phase are close to zero, which prevents us from use of the standard stationary phase method \cite{Fed}. 
We assume that $(f^{--}(p_1))''$ and $(f^{--}(p_2))''$ are distinct from zero. 
Then the contributions of the points $p_1$ and $p_2$ to the integral $I_1^{--}$ are described by
\begin{equation}\label{mm_p1}
    (I_1^{--})_{1} = \frac{e^{i\left(\frac{2}{3} \widetilde{\nu}^\frac{3}{2} -\sigma(\mathfrak{t} - \nu) - \frac{2}{3}\sigma^3\right)-i\frac{\pi}{4}}}{2\widetilde{\nu}^{\frac{1}{4}}} \left(1 + O\left(\frac{\sigma^2}{\widetilde{\nu}\left(\sigma- \sqrt{\widetilde{\nu}}\right)^3 } \right)\right)
\end{equation} 
and
\begin{equation}\label{mm_p2}
    (I_1^{--})_{2} = \frac{e^{-i\left(\frac{2}{3} \widetilde{\nu}^\frac{3}{2} + \sigma(\mathfrak{t} - \nu) + \frac{2}{3}\sigma^3\right)+i\frac{\pi}{4}}}{2\widetilde{\nu}^{\frac{1}{4}}} \left(1 + O\left(\frac{\sigma^2}{\widetilde{\nu}\left(\sigma+ \sqrt{\widetilde{\nu}}\right)^3} \right)\right).
\end{equation}
It is easy to see that the phases of exponentials in \eqref{mm_p1} and \eqref{mm_p2} agree with the expressions for eikonals \eqref{ph1} and \eqref{ph2}, respectively. 
Therefore, the contribution of $p_2$ corresponds to the contribution to the wavefield of a ray $\ell_2$ before passing the caustic, and the contribution of $p_1$ (since it is the critical point of the integral $I_1^{--}$ at $\nu>t$) corresponds to the contribution of a ray $\ell_2$ after passing the caustic. 
The expression $\widetilde{\nu}^{1/4}$ in the denominators in \eqref{mm_p1} and \eqref{mm_p2} is proportional to the square root of geometrical spreading $J$ of the rays $\ell_2$, see \cite{BabKir}.
It follows from \eqref{dist_c} and \eqref{kap} that $J = \sqrt{2\widetilde{n}/a}$.

The correction terms in \eqref{mm_p1} and \eqref{mm_p2} are small when the conditions 
\begin{equation*}
    \left|\sigma\mp \sqrt{\widetilde{\nu}}\right|^3\widetilde{\nu} \gg \sigma^2,
\end{equation*}
are met, or, in terms of the angle of inclination $\alpha$ of a ray and the distance to the caustic \eqref{dist_c}:
\begin{equation}\label{att2}
    \frac{k \widetilde{n} a^2|\alpha|^3}{x^2} \gg 1.
\end{equation}
It is clear from inequality \eqref{att2} that the formulas obtained above are not suitable for describing contributions to the wavefield of the rays close to the horizontal ray $l_B$ and in the vicinity of the caustic.

Therefore, the wave $u_2$, before and after the corresponding rays passing the caustic, is described by expressions \eqref{mm_p1} and \eqref{mm_p2}, respectively.

\textbf{2.} Proceed to the case when critical points $p_1$ and $p_2$ of the phase lie far from each other, but one of them is close to the end point $p=0$.
Remind that both critical points can merge the end point.

Consider merging of the point $p_1$ with the end point, which occurs when $\sigma>\sqrt{\mathfrak{t}}$ and the value of $\varepsilon = \sqrt{\mathfrak{t}}-\nu/2\sigma$ is small, see \eqref{eps}.
We expand the phase in a small neighborhood of $p=0$ up to quadratic terms:
\begin{equation*}
    f^{--}(p) = -\frac{2}{3}\mathfrak{t}^{\frac{3}{2}} + \frac{\nu^2}{4\sigma} + \varepsilon p - \frac{1}{4}\left(\frac{1}{\sqrt{\mathfrak{t}}} - \frac{1}{\sigma} \right)p^2 + O\left(\frac{p^3}{\mathfrak{t}^{\frac{3}{2}}}\right).
\end{equation*}
Omitting terms of the third order, extending integration from a small neighborhood of $p=0$ to the half-line $[0,\infty)$ and passing to the variable
\begin{equation*}
    q = \sqrt{\frac{1}{\sqrt{\mathfrak{t}}} - \frac{1}{\sigma}}\, \frac{p}{2},
\end{equation*}
we get
\begin{multline}\label{mm_01}
    (I_1^{--})_{01} = \frac{e^{-i\frac{2}{3}\mathfrak{t}^{\frac{3}{2}} + i\frac{\nu^2}{4\sigma}}}{2\sqrt{\pi}\sqrt{\sigma- \sqrt{\mathfrak{t}}}} \int_0^{\infty} e^{i\left(2\zeta_> q - q^2 \right)} \left[1 + O\left(\left(\frac{\sigma}{\sqrt{\mathfrak{t}} (\sigma-\sqrt{\mathfrak{t}})}\right)^{\frac{3}{2}} q^3\right) \right] \\
    = \frac{e^{i\left(-\frac{2}{3}\mathfrak{t}^{\frac{3}{2}} + \frac{\nu^2}{4\sigma} + \zeta_>^2\right) + i\frac{\pi}{4}}}{2\sqrt{\sigma- \sqrt{\mathfrak{t}}}} \Phi(i\zeta_>) \left[1 + O\left(\left(\frac{2\sigma\varepsilon}{\sigma-\sqrt{\mathfrak{t}}}\right)^3 \right) + O\left(\left(\frac{\sigma}{\sqrt{\mathfrak{t}}(\sigma-\sqrt{\mathfrak{t}})}\right)^{\frac{3}{2}} \right) \right].
\end{multline}
Here, $\Phi$ is the Fresnel integral
\begin{equation}\label{Fre}
   \Phi(z) = \frac{e^{-i\pi/4}}{\sqrt{\pi}} \int_{-\infty}^{z} e^{it^2} dt,  
\end{equation}
and
\begin{equation}\label{z>}
    \zeta_> = \frac{\sqrt{\sigma}\mathfrak{t}^{\frac{1}{4}}}{\sqrt{\sigma- \sqrt{\mathfrak{t}}}}\varepsilon= \sqrt{\frac{a\gamma}{x-a\gamma}} \sqrt{\frac{kr}{2}}(\gamma-\varphi),
\end{equation}
see \eqref{eps}. 
The value $\zeta_>^2$ equals, up to the constant summand $-2\mathfrak{t}^{3/2}/3$, difference between phase \eqref{ph12_as} of the wave $u_2$ after the rays $\ell_2$ pass the caustic and phase \eqref{ph_d_as} of the cylindrical wave.
The Fresnel integral in \eqref{mm_01} has an imaginary argument, which is atypical of diffraction problems.

Merging of the point $p_2$ with the end point $p=0$ occurs when $\varepsilon\approx 0$, $\sigma<\sqrt{\mathfrak{t}}$, and is described by the formula analogous to \eqref{mm_01}: 
\begin{multline}\label{mm_02}
    (I_1^{--})_{02} = \frac{e^{i\left(-\frac{2}{3}\mathfrak{t}^{\frac{3}{2}} + i\frac{\nu^2}{4\sigma} - \zeta_<^2\right) + i\frac{\pi}{4}}}{2\sqrt{\sqrt{\mathfrak{t}} - \sigma}} \Phi(-\zeta_<) \\
    \times \left[1 + O\left(\left(\frac{\varepsilon}{\sqrt{\mathfrak{t}} - \sigma}\right)^3 \right) + O\left(\left(\frac{\sigma}{\sqrt{\mathfrak{t}}(\sqrt{\mathfrak{t}} - \sigma)}\right)^{\frac{3}{2}} \right) \right]. 
\end{multline} 
Here,
\begin{equation}\label{z<}
    \zeta_< = \frac{\sqrt{\sigma} \mathfrak{t}^{\frac{1}{4}}}{\sqrt{\sqrt{\mathfrak{t}} - \sigma}} \varepsilon= \sqrt{\frac{a\gamma}{a\gamma - x}} \sqrt{\frac{kr}{2}}(\gamma-\varphi).
\end{equation}
The value $\zeta_<^2$ is equal, up to the constant summand $-2\mathfrak{t}^{3/2}/3$, to difference between the phase of the wave $u_2$ before the rays $\ell_2$ pass the caustic and the phase of the cylindrical wave, see \eqref{ph_d_as} and \eqref{ph12_as}.

It is not difficult to observe that 
\begin{equation}\label{JJ}
    \sqrt{|x-a\gamma|/a\gamma} = J(M)/J(O),
\end{equation}
where $J(M)$ and $J(O)$ are the values of geometrical spreading of the rays $\ell_2$ at the points $M=(x, \gamma x)$ and $O=(0,0)$ lying on the limit ray $l_O$ \eqref{lO1}.

Note that expressions \eqref{mm_01} and \eqref{mm_02} are singular at the point $Q$ where the caustic touches the limit ray $l_O$, cause $\sigma=\sqrt{\mathfrak{t}}$ for this point.
The correction terms in \eqref{mm_01} and \eqref{mm_02} are small under conditions
\begin{equation*}
    \sqrt{\mathfrak{t}}|\sigma-\sqrt{\mathfrak{t}}|\gg \sigma, \quad  \sigma\varepsilon\ll |\sigma-\sqrt{\mathfrak{t}}|,
\end{equation*}
which can be rewritten as follows
\begin{equation}\label{att3g}
    \mathfrak{m} \gamma |x - x_Q| \gg x, \quad  \mathfrak{m} x (\gamma - \varphi) \ll |x - x_Q|,
\end{equation}
see \eqref{gamma}, \eqref{Qxy}, \eqref{epsg}.
The first inequality implies that the observation point $M$ is positioned not too close to the point $Q$ and the second one characterizes the width of the transition zone surrounding $l_O$.

\textbf{3.} Now consider the case when critical points of phase $p_1$ and $p_2$ are close to each other, but distinct from the end point $p=0$.
This corresponds to the observation points lying near the caustic but far enough from the limit ray: $\widetilde{\nu} \approx 0$, however $\varepsilon$ is not small, see \eqref{kap} and \eqref{eps}.

Contribution of the point $p=0$ is calculated above and is given by formula \eqref{mm_0l}, applicable not too close to the limit ray. 

As $\widetilde{\nu} \approx 0$, the points $p_1$ and $p_2$ lie near the point $p_0 = \mathfrak{t} - \sigma^2$, which is a zero of the second derivative of \eqref{sec}.
The points $p_1$ and $p_2$ lie inside the integration interval $L_1$, see \eqref{L1}, when the inequality 
\begin{equation}\label{sgg1}
    \sigma\gg 1
\end{equation}
holds true.
We expand the phase $f^{--}(p)$ in a small neighborhood of $p_0$ up to cubic terms:
\begin{equation}\label{np0}
    f^{--}(p) = -\frac{2\sigma^3}{3} - \sigma(\mathfrak{t} - \nu) + \frac{\widetilde{\nu}^2}{4\sigma} + \frac{\widetilde{\nu}}{2\sigma}(p - p_0) - \frac{1}{24\sigma^3}(p - p_0)^3 + O\left(\frac{(p - p_0)^4}{\sigma^5}\right).
\end{equation}
Omitting terms of the fourth order, extending integration from a small neighborhood of $p_0$ to the whole axis and passing to the integration variable $q = (p - p_0)/2\sigma$, we obtain that the contribution of the critical points of phase to the integral $I_1^{--}$ is expressed in terms of the Airy function $v$:
\begin{multline}\label{mm_12}
    (I_1^{--})_{12} = \frac{e^{-i\left(\frac{2\sigma^3}{3} + \sigma(\mathfrak{t} - \nu) - \frac{\widetilde{\nu}^2}{4\sigma}\right)}}{2\sqrt{\pi}} \int_{-\infty}^{\infty} e^{i\left(\widetilde{\nu} q - \frac{q^3}{3}\right)} \left(1 + O\left(\frac{q + q^4}{\sigma}\right) \right) dq \\
    = e^{-i\left(\frac{2\sigma^3}{3} + \sigma(\mathfrak{t} - \nu) \right)} v(-\widetilde{\nu}) \left(1 + O\left(\frac{1+\widetilde{\nu}^2}{\sigma}\right) \right).
\end{multline}
Approximation \eqref{mm_12} holds in the area where 
\begin{equation}\label{att4}
    \widetilde{\nu}^2 \ll \sigma\quad \Leftrightarrow \quad k \widetilde{n}^2 \ll x,
\end{equation}
see \eqref{dist_c}.
Due to inequality \eqref{sgg1}, the argument of the Airy function in \eqref{mm_12} can be large in area \eqref{att4}. 

\textbf{4.} Finally, we proceed to the case when the critical points of phase $p_1$ and $p_2$ and the end point $p=0$ are all close to each other.
This corresponds to the observation points lying near the point $Q$ where the caustic touches the limit ray: $\sigma\approx \sqrt{\mathfrak{t}}$ and $\nu \approx 2\mathfrak{t}$.

As above, we expand the phase $f^{--}(p)$ in a neighborhood of $p_0 = t -\sigma^2$ (which is small now) up to cubic terms, see \eqref{np0}. 
We move terms of the fourth order to the amplitude, extend the integration to the half-line $[0,\infty)$ and pass to the variable $q = (p-p_0)/2\sigma$. 
Keeping in mind that $\sqrt{\mathfrak{t}} - \sigma$ is small, we come up with
\begin{equation}\label{mm_012}
    I_1^{--} = -e^{-i\left(\frac{2\sigma^3}{3} + \sigma(\mathfrak{t} - \nu) \right)} \mathcal{I}\left(-\widetilde{\nu}, \sqrt{\mathfrak{t}} - \sigma\right) \left(1 + O\left(\frac{(\sqrt{\mathfrak{t}}-\sigma)^2}{\sqrt{\mathfrak{t}}} \right) + O\left(\frac{\widetilde{\nu}^2 + 1}{\sqrt{\mathfrak{t}}} \right) \right).
\end{equation}
Here, $\mathcal{I}$ is the \textit{incomplete Airy function} (a special function of $B_3$-type catastrophe in terminology of \cite{Kryk,KrykLuk}):
\begin{equation*}
    \mathcal{I}(\eta, \xi) = \int_{\xi}^{-\infty} e^{i\left(\eta q + \frac{q^3}{3}\right)} dq.
\end{equation*}
The incomplete Airy functions emerge in many diffraction problems, see, e.g., \cite{Foc} and \cite{LevFel}.
Their generalizations, called \textit{multidimensional incomplete Airy functions} ($\mathcal{I}= \mathcal{I}(\eta_1, \eta_2, \ldots, \eta_n, \xi)$), were introduced in \cite{Blei}, where asymptotic expansions of integrals involving many arbitrarily positioned critical points studied. 
As far as we know, multidimensional incomplete Airy functions in diffraction problems first appeared in the work of Yu. I. Orlov \cite{Orl}. 

The asymptotics \eqref{mm_012} is applicable in the area where inequalities \eqref{att4} and
\begin{equation*}
    (\sigma-\sqrt{\mathfrak{t}})^2 \ll \sqrt{\mathfrak{t}} \quad \Leftrightarrow \quad \mathfrak{m} \frac{(x-x_Q)^2}{a^2} \ll \gamma,
\end{equation*}
cf. \eqref{att3g}, are satisfied, which characterize proximity of the observation point $M$ to the point $Q$.

It is clear that the approximations derived above match with each other in the intersections of their applicability areas.

\subsubsection{Integrals $I^{++}$, $I^{+-}$ и $I^{-+}$}
Now we discuss the results of asymptotic analysis of the rest integrals.

\textbf{1.} The asymptotics of the integral $I^{++}$, far from the limit ray, is given by the sum of contributions of the end point $p=0$ and the critical point of phase $p_3$.
The contribution of $p=0$ is similar to \eqref{mm_0l}:
\begin{equation}\label{pp_0l}
    (I_1^{++})_0 = i\frac{e^{i\frac{2}{3}\mathfrak{t}^{\frac{3}{2}} + i\frac{\nu^2}{4\sigma}}}{4\sqrt{\pi \sigma} \mathfrak{t}^{\frac{1}{4}}} \Bigg[\frac{i}{\varepsilon} \left(1 - \frac{5i}{24 \mathfrak{t}^{\frac{3}{2}}} + O\left(\frac{1}{\mathfrak{t}^3}\right)\right)  + \frac{1}{4\mathfrak{t}\,\varepsilon^2} + \left(\frac{1}{\sigma} + \frac{1}{\sqrt{\mathfrak{t}}}\right) \frac{1}{2 \varepsilon^3} \left(1 +\delta \right) \Bigg],
\end{equation}
see \eqref{eps} and \eqref{de}.
The contribution of $p_3$ has a form
\begin{equation}\label{pp_3}
    (I^{++})_3 = \frac{e^{i\left(\frac{2}{3} (\widetilde{\nu} + 2\nu)^\frac{3}{2} -\sigma(\mathfrak{t} + \nu) - \frac{2}{3}\sigma^3\right)-i\frac{\pi}{4}}}{2(\widetilde{\nu} + 2\nu)^{\frac{1}{4}}} \left(1 + O\left(\frac{\sigma^2}{(\sqrt{\widetilde{\nu} + 2\nu} - \sigma)^3(\widetilde{\nu} + 2\nu)} \right)\right).
\end{equation}
Comparing the phase of exponential in \eqref{pp_3} with expression \eqref{ph3}, we conclude that formula \eqref{pp_3} describes the wave $u_1$, corresponding to the ray family $\ell_1$, after reflection from the straight part of the boundary $S_+$. 
Geometrical spreading of these rays is proportional to the square root of the distance from the point $M'$ symmetric to the observation point $M$ relative to $S_+$, to the caustic $C$, see \eqref{dist_c+}.
The approximation \eqref{pp_0l}, \eqref{pp_3} is applicable in the area where conditions \eqref{att1} and
\begin{equation*}
    \mathfrak{t} + \nu \gg \sigma \quad \Leftrightarrow \quad \mathfrak{m} (b + y) \gg x
\end{equation*}
hold true.

Near the limit ray, the integral $I^{++}$ expressed in terms of the Fresnel integral \eqref{Fre}:
\begin{equation}\label{pp_03}
    I^{++} = \frac{e^{i\left(\frac{2}{3}\mathfrak{t}^{\frac{3}{2}} + \frac{\nu^2}{4\sigma} - \zeta_*^2 \right)-i\frac{\pi}{4}}}{2\sqrt{\sqrt{\mathfrak{t}} + \sigma}} \Phi(\zeta_*) \left[1 + O\left(\left(\frac{\sigma\varepsilon}{\sqrt{\mathfrak{t}} + \sigma}\right)^3 \right) \right],
\end{equation}
where
\begin{equation}\label{z*}
    \zeta_* = \frac{\sqrt{\sigma}\mathfrak{t}^{\frac{1}{4}} }{\sqrt{\sqrt{\mathfrak{t}} + \sigma}}\varepsilon= \sqrt{\frac{a\gamma}{x + a\gamma}} \sqrt{\frac{kr}{2}}(\gamma-\varphi), 
\end{equation}
cf. \eqref{z>} and \eqref{z<}. 
The quantity $\zeta_*^2$ is equal, up to the constant summand $2\mathfrak{t}^{3/2}/3$, to the difference between phase \eqref{ph3_as} of the wave $u_1$ after reflection from the straight part of the boundary and phase \eqref{ph_d_as} of the cylindrical wave. 
It is easy to check that $\sqrt{(x+a\gamma)/a\gamma} = J(M)/J(O)$, where $J(M)$ and $J(O)$ are the values of geometrical spreading of the rays $\ell_1$ at the points $M = (x, \gamma x)$ and $O = (0,0)$, respectively, cf. \eqref{JJ}.

\textbf{2.} The asymptotics of the integral $I^{+-}$ for $\nu>\mathfrak{t}$ is given by the contribution of the end point $p=0$, which has the form
\begin{multline}\label{pm_0l}
    (I_1^{+-})_0 = i\frac{e^{i\frac{2}{3}\mathfrak{t}^{\frac{3}{2}} + i\frac{\nu^2}{4\sigma}}}{4\sqrt{\pi \sigma} \mathfrak{t}^{\frac{1}{4}}} \Bigg[\frac{i}{\tilde{\varepsilon}} \left(1 - \frac{5i}{24 \mathfrak{t}^{\frac{3}{2}}} + O\left(\frac{1}{\mathfrak{t}^3}\right)\right)  + \frac{1}{4\mathfrak{t}\,\tilde{\varepsilon}^2} \\
    + \left(\frac{1}{\sigma} + \frac{1}{\sqrt{\mathfrak{t}}}\right) \frac{1}{2 \tilde{\varepsilon}^3} \left(1 +O\left(\frac{1}{\mathfrak{t}^{\frac{3}{2}}}\right) + O\left(\frac{1}{\mathfrak{t}\sigma}\right) \right) \Bigg].
\end{multline} 
Here, we introduce the notation
\begin{equation}\label{tileps}
    \tilde{\varepsilon} = \sqrt{\mathfrak{t}} + \frac{\nu}{2\sigma} = 2\sqrt{\mathfrak{t}} - \varepsilon.
\end{equation}

The asymptotics of  $I^{+-}$ for $\nu < \mathfrak{t}$ is described by the sum of end point contribution \eqref{pm_0l} and the contribution of the critical point of phase $p_1$.
The latter is given by
\begin{equation}\label{pm_1}
    (I^{+-})_1 = \frac{e^{i\left(\frac{2}{3} \widetilde{\nu}^\frac{3}{2} -\sigma(\mathfrak{t} - \nu) - \frac{2}{3}\sigma^3\right)-i\frac{\pi}{4}}}{2\widetilde{\nu}^{\frac{1}{4}}} \left(1 + O\left(\frac{\sigma^2}{\widetilde{\nu}(\widetilde{\nu} - \sigma)^3} \right)\right),
\end{equation}
cf. \eqref{mm_p1} and \eqref{mm_p2}. 
Formula \eqref{pm_1} describes the rays $\ell_1$ before they reflect from the straight part of the boundary, see \eqref{ph1}.
Approximation \eqref{pm_1} is applicable in the area where inequality \eqref{att2} is satisfied.

\textbf{3.} The critical point $p_4$ of phase of the integral $I^{-+}$ is negative and lies far from the end point $p=0$ for all values of $\sigma$ and $\nu$.
Thus, the asymptotics of $I^{-+}$ is given by the end point $p=0$ contribution:
\begin{multline}\label{mp_0l}
    (I_1^{-+})_0 = \frac{e^{-i\frac{2}{3}\mathfrak{t}^{\frac{3}{2}} + i\frac{\nu^2}{4\sigma}}}{4\sqrt{\pi \sigma} \mathfrak{t}^{\frac{1}{4}}} \Bigg[\frac{i}{\tilde{\varepsilon}}\left(1 + \frac{5i}{24 \mathfrak{t}^{\frac{3}{2}}} + O\left(\frac{1}{\mathfrak{t}^3}\right)\right) - \frac{1}{4\mathfrak{t} \, \tilde{\varepsilon}^2}\\
    + \left(\frac{1}{\sigma} - \frac{1}{\sqrt{\mathfrak{t}}}\right) \frac{1}{2 \tilde{\varepsilon}^3} \left(1 + O\left(\frac{1}{\mathfrak{t}^{\frac{3}{2}}}\right) + O\left(\frac{1}{\mathfrak{t}\sigma}\right) \right) \Bigg]
\end{multline}

\textbf{4.} Above, the geometrical interpretation of contributions of the critical points in the asymptotics of integrals has been established.
Now consider contributions of the end point $p=0$.

The sum of contributions of the end point $p=0$ to the integrals $I_1^{--}$ and $I_1^{++}$, see \eqref{mm_0l} and \eqref{pp_0l}, has a form
\begin{multline}\label{mmpp0}
    (I_1^{--})_0 + (I_1^{++})_0 = -\frac{e^{i\frac{\nu^2}{4\sigma} - i\frac{\pi}{4}}}{2\sqrt{\pi \sigma} \mathfrak{t}^{\frac{1}{4}}}\frac{\cos\left(\frac{2}{3}\mathfrak{t}^{\frac{3}{2}} + \frac{\pi}{4}\right)}{\varepsilon} \left(1 + \delta \right) \\
    -\frac{e^{i\frac{\nu^2}{4\sigma} - i\frac{\pi}{4}}}{2\sqrt{\pi \sigma} \mathfrak{t}^{\frac{1}{4}}}\frac{\sin\left(\frac{2}{3}\mathfrak{t}^{\frac{3}{2}} + \frac{\pi}{4} \right)}{\varepsilon} \left[\frac{5}{24\mathfrak{t}^{\frac{3}{2}}} + \frac{1}{4\mathfrak{t} \varepsilon} + \frac{1}{2\sqrt{\mathfrak{t}}\, \varepsilon^2} \left(1 + \delta \right) \right].
\end{multline}
see \eqref{eps} and \eqref{de}.
Deriving the asymptotics for $v'(z)$ as $z\to -\infty$ from \eqref{vas1}, substituting $z=-\mathfrak{t}$ and taking into account that $v'(\mathfrak{t})=0$, we obtain following relation:
\begin{equation}\label{c-s}
    \cos\left(\frac{2}{3}\mathfrak{t}^{\frac{3}{2}} + \frac{\pi}{4}\right) = \frac{1}{24 \mathfrak{t}^{\frac{3}{2}}}\sin\left(\frac{2}{3}\mathfrak{t}^{\frac{3}{2}} + \frac{\pi}{4}\right)\left(1 + O\left(\frac{1}{\mathfrak{t}^3}\right)\right)
\end{equation}
Using \eqref{c-s}, we simplify expression \eqref{mmpp0}:
\begin{equation}\label{mmpp0+}
    (I_1^{--})_0 + (I_1^{++})_0 \approx 
    -\frac{e^{i\frac{\nu^2}{4\sigma} - i\frac{\pi}{4}}}{2\sqrt{\pi \sigma} \mathfrak{t}^{\frac{1}{4}}}\frac{\sin\left(\frac{2}{3}\mathfrak{t}^{\frac{3}{2}} + \frac{\pi}{4} \right)}{\varepsilon} \left[\frac{1}{4\mathfrak{t}^\frac{3}{2}} + \frac{1}{4\mathfrak{t} \varepsilon} + \frac{1}{2\sqrt{\mathfrak{t}}\, \varepsilon^2}  \right].
\end{equation}
The sum of end point contributions in $I_1^{+-}$ and $I_1^{-+}$, see \eqref{pm_0l} and \eqref{mp_0l}, differs from \eqref{mmpp0+} only by replacing $\varepsilon$ with $\tilde{\varepsilon}$ \eqref{tileps}:
\begin{equation}\label{pmmp0}
    (I_1^{+-})_0 + (I_1^{-+})_0 \approx 
    -\frac{e^{i\frac{\nu^2}{4\sigma} - i\frac{\pi}{4}}}{2\sqrt{\pi \sigma} \mathfrak{t}^{\frac{1}{4}}}\frac{\sin\left(\frac{2}{3}\mathfrak{t}^{\frac{3}{2}} + \frac{\pi}{4} \right)}{\tilde{\varepsilon}} \left[\frac{1}{4\mathfrak{t}^\frac{3}{2}} + \frac{1}{4\mathfrak{t} \tilde{\varepsilon}} + \frac{1}{2\sqrt{\mathfrak{t}}\, \tilde{\varepsilon}^2}  \right].
\end{equation}
Finally, summing up \eqref{mmpp0+} and \eqref{pmmp0}, we get
\begin{equation}\label{I10}
    (I_1)_0 \approx -\frac{e^{i\frac{\nu^2}{4\sigma} - i\frac{\pi}{4}}}{\sqrt{\pi \sigma}}\frac{\sin\left(\frac{2}{3}\mathfrak{t}^{\frac{3}{2}} + \frac{\pi}{4} \right)}{\mathfrak{t}^{\frac{1}{4}}} \frac{t + (\nu/2\sigma)^2}{(t - (\nu/2\sigma)^2)^3}.
\end{equation}

It is clear from \eqref{ph_d} that the phase of exponential in \eqref{I10} agree with the eikonal of cylindrical wave in the area where
\begin{equation}\label{phi4}
    kr\varphi^4 \ll 1,
\end{equation}
see \eqref{ph_d}.
We rewrite expression \eqref{I10} in polar coordinates \eqref{pol} using \eqref{xystr} and \eqref{gamma}. 
Matching it with ray formula \eqref{cyl} under conditions \eqref{kr} and \eqref{phi4}, we obtain that the diffraction coefficient has the form 
\begin{equation}\label{A}
    A(\varphi; k) \approx 2 \sqrt{\frac{2}{\pi}} \frac{\sin\left(\frac{2}{3}\mathfrak{t}^{\frac{3}{2}} + \frac{\pi}{4} \right)}{ka \mathfrak{t}^{\frac{1}{4}}} \frac{\gamma^2 + \varphi^2}{(\varphi^2 - \gamma^2)^3}  e^{- i\frac{\pi}{4}} \approx 2 \sqrt{\frac{2}{\pi}} \frac{v(-\mathfrak{t})}{ka} \frac{\gamma^2 + \varphi^2}{(\varphi^2 - \gamma^2)^3}  e^{- i\frac{\pi}{4}},
\end{equation}
see \eqref{vas1}.
The expression \eqref{A} is applicable in the area where inequality \eqref{att1} holds and the polar angle is small, $\varphi \ll 1$.
The diffraction coefficient \eqref{A} is linear in the jump of curvature $1/a$, as it is in other problems of diffraction by jump of curvature (see, e.g., \cite{Pop71,ZKAkZh,ZKWM23,ZloKisWM}). 

In the area where $\varphi \gg \gamma$ the expression \eqref{A} becomes
\begin{equation}\label{A>}
    A(\varphi; k) \approx \sqrt{\frac{2}{\pi}} \frac{v(-\mathfrak{t})}{ka} \frac{2}{\varphi^4}  e^{- i\frac{\pi}{4}}.
\end{equation}
The formula \eqref{A>} coincides with the diffraction coefficient derived in \cite{ZKAkZh}, where diffraction of small-number ($\mathfrak{t} =O(1)$) whispering gallery mode was considered.

Moreover, formula \eqref{A} agrees with the expression for diffraction coefficient obtained in \cite{Zlo} in the case of \textit{non-tangential} incidence. 
Indeed, for a plane incident wave and the Neumann boundary condition the diffraction coefficient $A_{nt}$ has a form (see \cite{Zlo})
\begin{equation}\label{Ant}
    A_{nt}(\varphi, k) = \sqrt{\frac{2}{\pi}} \frac{1}{ka} \frac{\cos\gamma\cos\varphi - 1}{(\cos\varphi - \cos\gamma)^3} e^{-i\frac{\pi}{4}}.
\end{equation}
Here, $\gamma$ is the grazing angle of incident wave, $1/a$ is the amplitude of curvature jump and $\varphi$ is the polar angle. 
Obviously, for small $\gamma$ and $\varphi$ expression \eqref{Ant} matches with \eqref{A} up to the factor $v(-\mathfrak{t})/2$. 

\subsection{Investigation of integral $I_2$}
Now we turn to the study of the integral $I_2$ over the interval $L_2$, see \eqref{L2}, where the Airy function changes slowly.
For convenience, the integral is represented as a sum
\begin{equation*}
    I_2 = I_2^- + I_2^+,
\end{equation*}
where
\begin{equation}\label{pm}
    I_2^\pm=\frac{e^{-i\pi/4}}{2\sqrt{\pi \sigma}} \int_{L_2} v(p -\mathfrak{t}) e^{i\frac{(p\pm\nu)^2}{4\sigma}} dp.
\end{equation}

Consider the area where $\sigma\ll 1$, then $1/4\sigma$ in \eqref{pm} is a large parameter. 
The integrand of $I_2^+$ has no critical points on $L_2$, therefore the integral is of order of $\sqrt{\sigma}$.
As $\nu \approx \mathfrak{t}$, the phase of the integral $I_2^-$ has a critical point $p = \nu$, and its asymptotics is given by
\begin{equation}\label{2m}
    I_2^- = v(\nu - \mathfrak{t}) \left(1 + O(\sigma(\nu - \mathfrak{t}))\right).
\end{equation}
Formula \eqref{2m} describes the penetration of the incident whispering gallery mode to the right of the $y$ axis.
If $|\nu - \mathfrak{t}| \gg 1$, then the integral $I_2^-$ has no critical points and does not allow asymptotic simplification. 

In the area where $\sigma\gtrsim 1$, integrals \eqref{pm} also can not be simplified. 

Recall that the integral $I_3$ over the interval $L_3$ \eqref{L3} where the Airy function decreases exponentially, is negligible compared to $I_2$.

\section{Conclusions}

Diffraction of a large-number whispering gallery mode by jumply straitening of curvature has been considered.
The wavefield has been described in detail in a small neighborhood (see \eqref{obl}) of the jump point $O$ to the right of it within the framework of the parabolic equation method. 
A comprehensive geometrical analysis of the problem has been carried out, which allows to give a geometrical interpretation to the derived asymptotic formulas. 
Let us summarise results obtained above and compare them with the ones of the study \cite{ZKAkZh} devoted to the case of a small-number mode.

\begin{figure}[H]
    \noindent\centering{\includegraphics[width=0.45\textwidth]{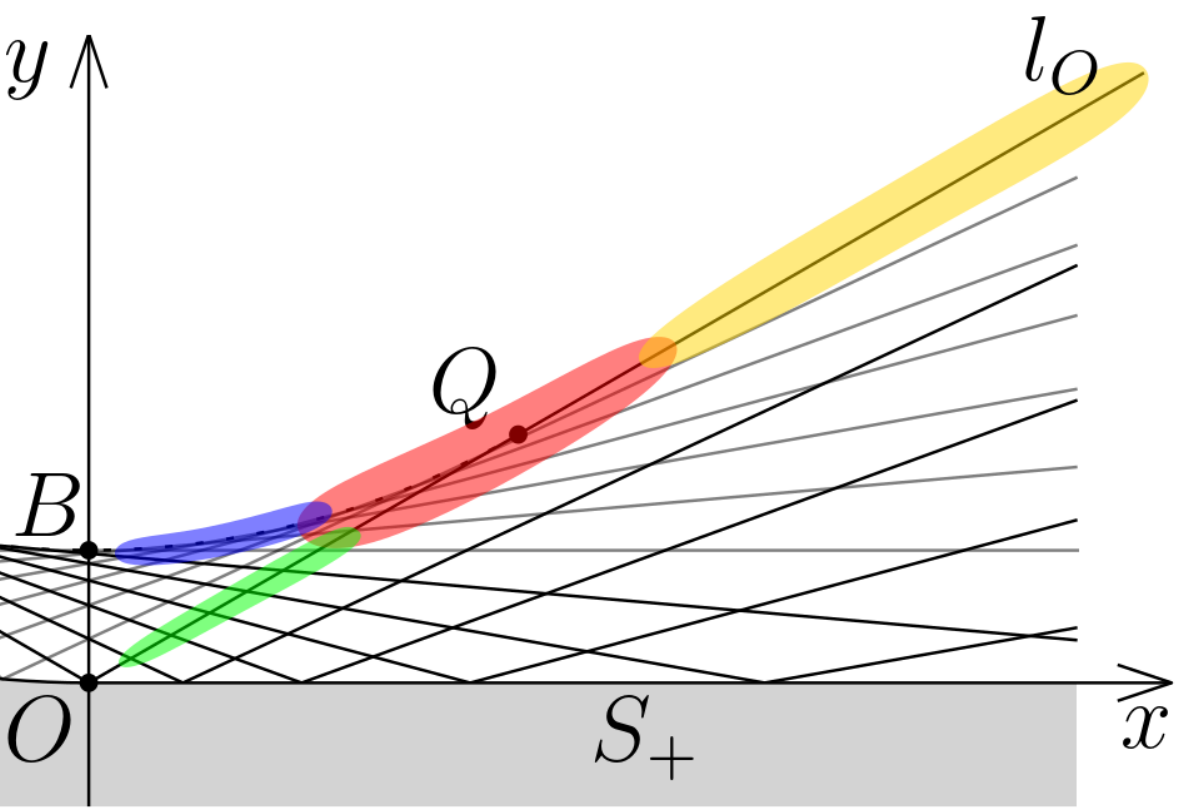}}
    \caption{Sketch of rays and boundary layers  (diffracted wave is omitted)}\label{f5}
\end{figure}

A sketch of boundary layers for a large-number mode incidence is shown in Fig. \ref{f5}. 
Where the layers overlap, the corresponding asymptotics match.  
Note that the boundary layers surrounding the limit ray (green and yellow zones) and the caustic (blue zone) intersect with the boundary layer around the point $Q$ (red zone), but not with each other. 

Both for moderate and large $\mathfrak{t}$ (see \eqref{wgas}), the cylindrical diffracted wave diverging from $O$ arises. 
Also, the whispering gallery mode penetrates to the right of the $y$ axis, which process is described by the formula \eqref{2m}. 
In other respects, the structure of the wavefield in the case of $\mathfrak{t}\gg 1$ significantly differs from the one in the case of $\mathfrak{t} = O(1)$. 

For large $\mathfrak{t}$, terms \eqref{mm_p1}, \eqref{mm_p2}, \eqref{pp_3} corresponding to the rays are clearly distinguishable in the asymptotics of the field.\footnote{Note that contributions of the rays close to the horizontal ray $l_B$ (see Fig. \ref{f5}), as mentioned in section 5.1 and especially around \eqref{att2}, can not be described analytically. 
Apparently, this is because of the insufficient accuracy of approximation \eqref{wgas} for the incident wave. 
}
To the right of $O$, a rather extended caustic emerges.  
In its vicinity (blue zone in Fig. \ref{f5}) the wavefield is described by the sum of expression \eqref{mm_12} involving the Airy function and diffracted wave \eqref{cyl}, \eqref{A}. 
Noteworthy is that in the area where $\varphi \gg \gamma$ diffraction coefficient \eqref{A} coincides with the one for $\mathfrak{t} = O(1)$.
Moreover, formula \eqref{A} agrees with the expression for diffraction coefficient for \textit{non-tangential} incidence.

In the case of moderate $\mathfrak{t}$, the diffracted wave has a singularity on the straight part of the boundary.
In its vicinity, the wavefield is described by the parabolic cylinder functions $D_{-4}$ with arguments independent of the geometrical parameters of the problem, see \cite{ZKAkZh}.
However, in the case of large $\mathfrak{t}$, diffracted wave \eqref{cyl}, \eqref{A} is singular on the limit ray $l_O$, but not on $S_+$, see Fig. \ref{f5}.
Transition zone surrounding $l_O$ is divided into two parts by the point $Q$.  
To the left of $Q$ (green zone in Fig.~\ref{f5}), the wavefield is expressed by sum of the Fresnel integrals \eqref{pp_03} and \eqref{mm_02} and the background \eqref{pm_0l}, \eqref{mp_0l}. 
Formulas \eqref{pp_03} and \eqref{mm_02} describe merging of diffracted wave with the wave $u_1$ after its reflection from $S_+$ and with the wave $u_2$ before passing the caustic, respectively.
To the right of $Q$ (yellow zone in Fig. \ref{f5}), the asymptotics of wavefield is sum of the Fresnel integrals \eqref{pp_03} and \eqref{mm_01} and the background. 
Formula \eqref{mm_01} describes merging of diffracted wave with the wave $u_2$ after passing the caustic and has an imaginary argument, which is unusual for diffraction problems. 
Expressions \eqref{z*}, \eqref{z<} and \eqref{z>} in the arguments of the Fresnel integrals depend on curvature radius of $S_-$ and grazing angle $\gamma$ (see Fig. \ref{f2}).

Finally, in the boundary layer surrounding the point $Q$ (red zone in Fig. \ref{f5}) where the caustic touches the limit ray, the wavefield is described by the sum of incomplete Airy function \eqref{mm_012} and the background. 

The author is grateful to A.P. Kiselev for his help in preparing the article.

The work is supported by the RSF grant 22-11-00070.


\begin {thebibliography}{99}

\bibitem{Wes} Weston V.H., The effect of a discontinuity in curvature in high-frequency scattering, IRE T. AP 10 (6) (1962) 775--780.

\bibitem{Pop71} Popov A.V., Backscattering from a line of jump of curvature, in: Trudy V Vses. Sympos. Diffr. Raspr. Voln, Nauka, Leningrad, 1971, pp. 171--175. [in Russian]

\bibitem{KamKel} Kaminetzky L., Keller J.B., Diffraction coefficients for higher order edges and vertices, SIAM J. Appl. Math. 22 (1) (1972) 109--134.

\bibitem{RogKis} Rogoff Z.M., Kiselev A.P. Diffraction at jump of curvature on an impedance boundary, Wave Motion 33 (2) (2001) 183--208.

\bibitem{KirPhi95} Kirpichnikova N.Ya., Philippov V. B., Behavior of surface waves at transition through a junction line on the boundary of an
elastic homogeneous isotropic body, J. Math. Sci. 91 (1998) 2757--2767.  	

\bibitem{KirPhi97} Kirpichnikova N.Ya., Philippov V.B., Diffraction of the whispering gallery waves by a conjunction line, J. Math. Sci.
96 (1999) 3342--3350.

\bibitem{KirPhi98} Philippov V.B., Kirpichnikova N.Ya., The edge wave in the problem of diffraction on a boundary with a jump of curvature, J. Math. Sci. 102 (2000) 274--287.

\bibitem{ZloKisWM} Zlobina E.A,  Kiselev A.P., Boundary-layer approach to high-frequency diffraction by a jump of curvature, Wave Motion 96 (2020) 102571.

\bibitem{Zlo} Zlobina E.A., Short-wavelength diffraction by a contour with nonsmooth curvature. Boundary layer approach, J. Math. Sci. 277 (4) (2023), 586--597.

\bibitem{ZloKisRE} Zlobina E.A., Kiselev A.P., Transition zone in high-frequency diffraction on impedance contour with jumping curvature. Kirchhoff’s method and boundary
layer method, J. Comm. Tech. El. 67 (2) (2022) 130--139.

\bibitem{ZKWM23} Zlobina E.A., Kiselev A.P. The Malyuzhinets---Popov diffraction problem revisited, Wave Motion 121 (2023) 103172.

\bibitem{ZKAkZh} Zlobina E.A., Kiselev A.P., Diffraction of a whispering gallery mode at a jumply straightening of the boundary, Acoust. Phys. 69 (2023) 133--142. 

\bibitem{AA} Zlobina E.A., Kiselev A.P., Short wave diffraction on a contour with a H\"{o}lder singularity of the curvature, St. Petersburg Math. J., 33 (2) (2022) 207--222. 

\bibitem{BorKin} Borovikov V.A., Kinber B.E., Geometrical Theory of Diffraction, Institute of Electrical Engineers, London, 1994.

\bibitem{Foc} Fock V.A., Electromagnetic Diffraction and Propagation Problems, Pergamon Press, Oxford, 1965.

\bibitem{BabKir} Babich V.M., Kirpichnikova N.Y., The Boundary Layer Method in Diffraction Problems, Springer, Berlin, 1979.


\bibitem{LMMP} Lanin A.I., Popov M.M., Behaviour of the whispering gallery rays in a vicinity of a point where curvature of the boundary vanishes, J. Soviet Math. 20 (1) (1982) 1840--1845.

\bibitem{Berry} Berry M.V., Inflection reflection: images in mirrors whose curvature changes sign, Eur. J. Phys. 42 (2021) 065301.

\bibitem{BabBul} Babich V.M., Buldyrev V.S., Asymptotic Methods in Short-Wavelength Diffraction Theory, Alpha Science, Oxford, 2007.

\bibitem{Fed} Wong R., Asymptotic approximations of integrals, SIAM, Philadelphia, 1989.

\bibitem{Kryk} Kryukovsky A.S., Uniform asymptotic theory of edge and corner catastrophes, RosNOU, Moscow, 2013. [in Russian] 

\bibitem{KrykLuk} Kryukovsky A.S., Lukin D.S., Edge Catastrophes in Diffraction Problems, J. Commun. Technol. Electron. 64, (2019) 1224--1229.

\bibitem{LevFel} Levey L., Felsen L. B., On incomplete Airy functions and their application to diffraction problems, Radio Sci. 4 (10) (1969) 959--969.

\bibitem{Blei} Bleistein N., Uniform asymptotic expansions of integrals with many nearby stationary points and algebraic singularities, J. Math. Mech. 17 (6) (1967) 533--559.

\bibitem{Orl} Orlov Yu.I., Uniform asymptotics of the diffraction field on a curved wedge in the presence of caustics. Radiotehnika i Electronika 20 (2) (1975) 242--248. [in Russian]


\end{thebibliography}
\end{document}